\documentclass{article}
\usepackage[a4paper, total={7in, 9in}]{geometry}
\usepackage{natbib}
\usepackage{amsmath}
\setcitestyle{authoryear,open={(},close={)}}
\usepackage{graphicx} % Required for inserting images
\usepackage{amsmath}
\usepackage{xcolor}
\usepackage{authblk}

\title{An integration of decision trees into latent class modeling with covariates}
\author[1,3]{Johan Lyrvall}
\author[2]{Felix Clouth}
\affil[1]{National Institute for Research in Digital Science and Technology (Inria), France}
\affil[2]{Department of Methodology and Statistics, Tilburg University, The Netherlands}
\affil[3]{Corresponding author: johan.lyrvall@outlook.com}
\date{}

\begin{document}

\maketitle

\abstract{
\noindent
We propose a novel methodology for fitting decision trees to latent classes.
The latent class analysis methodological literature has previously been focusing on logistic models of class membership given covariates, which has important drawbacks in the presence of complex interactions between covariates: logistic models are easily misspecified by omitting some interaction terms and complexity of interpretation increases rapidly with inclusion of higher-order interaction terms.
Our proposed approach directly integrates decision tree modeling into the latent class analysis framework to model the covariate effects as an easily interpretable tree leading the eye through combinations of predictors to a final conditional classification.
The novel methodology does not require any non-traditional assumptions for latent class models with covariates, is based on well-established routines of model estimation, and can be readily implemented in existing software.
In the present paper, we focus on decision trees for binary covariates.
We present the proposed approach, describe two tree pruning strategies, and provide a real-data illustration.
Important extensions include generalizing the approach to multinomial and continuous covariates by developing more advanced within-predictor splitting procedures, and developing and evaluating alternative tree pruning strategies.
The ultimate aim of this work is to initiate a novel research line in the latent class analysis methodological literature, and to facilitate the advancement of applied research via a novel data analytical tool.
}

\section{Introduction}

Latent class analysis \citep{goodman1974} is a popular clustering approach, latent variable approach, and mixture modeling approach for categorical data.
This is a popular method for identifying an unobserved, or latent, classification of individuals based on multiple observed categorical indicators.
Examples of applications can be taken from diverse behavioral scholarships like economics \citep{angelinilyrvall2025}, political science \citep{oseretal2013}, and sociology \citep{vanreesvermuntverboord1999}.

In addition to identifying classes of individuals, latent class analysis typically involves analyzing external causes of class membership.
This involves the inclusion of covariates, or predictors in the latent class model.
Approaches to estimating latent class models with covariates are actively researched in latent class analysis methodological scholarship, with recent examples including \citet{bakkkuha2018,vermuntmagidson2021,lyrvalletal2024,lyrvallkuhaoser2025}.
This literature is concerned with (multinomial) logistic models of class membership given covariates.
While this is often a natural modeling choice, in some contexts, it exhibits common drawbacks shared by linear specifications generally.
In this paper, we address one particular drawback: complexity of interpretation due to complex interactions between the covariates.

More specifically, we propose a novel approach for modeling complexly interacting covariate effects by means of a decision tree.
Decision trees are popular in standard (i.e., not latent) classification for modeling complex interactions in the predictor space.
This is mechanized by means of recursive partitioning of the predictor space through binary splits.
The resulting structure is an easily interpretable tree leading the eye through combinations of predictors to a final conditional classification.
Applied economic research by \citet{galletta2016} exemplifies decision tree modeling in behavioral scholarship.
In the present paper, we focus on decision trees for binary covariates.

Our proposed approach is a stepwise latent-class tree approach that preserves the measurement model while recursively partitioning the covariate space according to differences in latent class prevalences.
As such, our approach directly integrates decision tree modeling into the latent class analysis framework.
The novel methodology does not require any non-traditional assumptions.
It is based on well-established routines of model estimation, and can be readily implemented in existing software.

Its closest methodological relative is the hybrid method by \citet{magidsonvermunt2005} combining the CHAID algorithm and latent class analysis.
The hybrid method involves computing posterior class membership probabilities which are then treated as an observed dependent variable when applying the CHAID algorithm with the covariates.
However, this is conflicting with what is now the benchmark way of specifying latent class models, namely taking the covariates as affecting class membership probability as opposed to posterior class membership probability.
The proposed approach is fully compatible with this benchmark specification.
This justifies our presentation of it as not a hybrid approach but an integration of decision tree modeling into the latent class analysis framework.

The present article is structured as follows.
In the next section, we present the methodological framework.
Subsequently, we describe our decision tree approach to latent class prediction.
In the section following that, we describe two alternative tree pruning strategies.
We then provide a real-data illustration of the proposed approach using data from the 2017-2018 National Health and Nutrition Examination Survey (NHANES; \citet{johnson2013}).
In the final section, we conclude.

\section{Methodological framework}

Let $X$ be an unobserved, or latent, discrete classification of $N$ individuals into $T \geq 2$ discrete types, or classes, in which individual $i = 1,\dots,N$ belongs to exactly one type.
We denote by $P(X=t)$ the unconditional probability of belonging to type $t=1,\dots,T$.
Let $\mathbf{Y} = (Y_h; h=1,\dots,H)$ be $H$ observed indicators of $X$, and $r_h = 1,\dots,R_h$ a particular response category for $Y_h$ (which is dichotomous if $R_h=2$ and polytomous if $R_h \geq 3$).
We denote by $P(\mathbf{Y}=\mathbf{r}) = P(Y_h=r_h; h=1,\dots,H)$ the unconditional joint probability of a particular response pattern across the $H$ indicators.
In standard latent class analysis \citep[e.g.][]{goodman1974}, the relationship between $X$ and $\mathbf{Y}$ is defined as

\begin{equation}\label{eq:standard}
    \begin{split}
        P(\mathbf{Y}=\mathbf{y}) & = \sum_{t=1}^T P(X=t) P(\mathbf{Y}=\mathbf{r} \vert X=t) \\
        & = \sum_{t=1}^T P(X=t) \left\{ \prod_{h=1}^H P(Y_h=r_h \vert X=t) \right\}.
    \end{split}
\end{equation}
\\
Here, $P(\mathbf{Y}=\mathbf{r} \vert X=t)$ is the conditional joint probability of response pattern $\mathbf{r}$ given class $t$, and $P(Y_h=r_h \vert X=t)$ the conditional probability of response $r_h$ to item $Y_h$ given class $t$.
As can be seen in the first line, the joint probability of the responses is a mixture of their class-specific joint probabilities, weighted by the class probabilities.
As can be seen in the second line, the responses are conditionally independent of each other given the classes, or locally independent of each other within the classes.

\eqref{eq:standard} can be extended to include external class predictors, or covariates.
Let $\mathbf{Z} = (Z_p; p=1,\dots,P)$ be $P$ covariates.
We note that, while the present paper focuses on binary covariates, the following exposition of the latent class model with covariates allows also for multinomial and continuous covariates.
We denote by $P(X=t \vert \mathbf{Z})$ the conditional probability of class $t$ given $\mathbf{Z}$.
The standard latent class model with covariates \citep[e.g.][]{vermunt2010,bakkkuha2018} is defined as

\begin{equation}\label{eq:covariates}
    P(\mathbf{Y}=\mathbf{r} \vert \mathbf{Z}) = \sum_{t=1}^T P(X=t \vert \mathbf{Z}) \left\{ \prod_{h=1}^H P(Y_h=r_h \vert X=t) \right\},
\end{equation}

where

\begin{equation}\label{eq:logistic}
    P(X=t \vert \mathbf{Z}) = \frac{\exp(\gamma_{t0}+\boldsymbol{\gamma}_t\mathbf{Z})}{1+\sum_{s=2}^T\exp(\gamma_{s0}+\boldsymbol{\gamma}_s\mathbf{Z})}.
\end{equation}
\\
As can be seen in \eqref{eq:covariates}, the indicators are conditionally independent of the covariates given the classes.
As can be seen in \eqref{eq:logistic}, the association between the classes and the covariates is a logistic regression with intercept $\gamma_{t0}$ and slopes $\boldsymbol{\gamma}_t = (\gamma_{tp};p=1,\dots,P)$.
Note that this flexible parameterization allows for absence of intercept via the possibility of $\gamma_{t0}=0$.

The $T \times \sum_{h=1}^H R_h$ parameters represented by $P(Y_h=r_h \vert X=t)$ is commonly referred to as the measurement model, and the $(P+1) \times T$ parameters represented by the right-hand side of \eqref{eq:logistic} as the structural model.
In applied research, the measurement model is used to define the classes, and the structural model to describe how common the classes are in the population and how this prevalence varies with the covariates.
The research interest of this paper lies in the structural model.

In applied latent class analysis, structural models are typically fitted with a linear specification.
However, structural models with complex interaction effects may be difficult to interpret on the basis of a linear specification.
Furthermore, when the interactions are unknown a priori, they may be difficult to discover in the first place.
The latent class analysis methodological literature has not yet addressed this problem.
In the next section, we propose an innovative modeling approach as a suggested solution.

\section{The proposed modeling approach with decision trees}

We propose an alternative modeling approach using decision trees to discover and describe complex interactions in the structural model.
The present work focuses on trees with binary covariates.
For simplicity of exposition, we introduce the following shorthand notation.

\textbf{Parameters:} Let $\boldsymbol{\phi} = (\phi_{hrt}; h=1,\dots,H; r=1,\dots,R_h; t=1,\dots,T)$ be the parameter space for the measurement model, with $\phi_{hrt} = P(Y_h=r_h \vert X=t)$; $\boldsymbol{\pi} = (\pi_t; t=1,\dots,T)$ the parameter space for the unconditional structural model in \eqref{eq:standard}, with $\pi_t = P(X=t)$; and $\boldsymbol{\pi}^p = (\pi_t^{p_v}; t=1,\dots,T; v=0,1)$ the parameter space for the structural model in \eqref{eq:covariates} when $\mathbf{Z} = Z_p$, with $\pi_t^{p_v} = P(X=t \vert Z_p=v)$.

\textbf{Samples:} Let $S$ be the full sample; $S^{p_v}$ the subsample for which value $v=0,1$ is observed on covariate $Z_p$; and $N^{p_v}$ the size of subsample $S^{p_v}$.

We present the proposed approach assuming that the number of latent classes $T$ has been selected prior to the inclusion of covariates in the latent class model.
Model selection is a separate and important task on which we do not focus here.
We refer readers interested in best practices for class enumeration to \citet{lyrvalloserbakk2026}.

The proposed approach can be described by means of the following algorithm:

\begin{enumerate}
    \item Fit \eqref{eq:standard} to the full sample $S$ by maximizing the log-likelihood
    \begin{equation*}
        \ell_1(\boldsymbol{\pi},\boldsymbol{\phi} \vert S) = \sum_{i=1}^N \log \left[ \sum_{t=1}^T P(X=t) \left\{ \prod_{h=1}^H P(Y_h=y_{ih} \vert X=t) \right\} \right],
    \end{equation*}
    to obtain estimates $\widehat{\boldsymbol{\pi}}$ and $\widehat{\boldsymbol{\phi}}$, and posteriors $\widehat{\boldsymbol{\pi}}_{\mathbf{y}}$.
    \item For each covariate $Z_p$, fit \eqref{eq:covariates} with $\mathbf{Z} = Z_p$ to $S$ conditionally on $\widehat{\boldsymbol{\phi}}$ from step 1 by maximizing the pseudo log-likelihood
    \begin{equation*}
        \ell_{2}(\boldsymbol{\pi}^p \vert \widehat{\boldsymbol{\phi}},S) = \sum_{i=1}^N \log \left[ \sum_{t=1}^T P(X=t \vert Z_p=v_i) \left\{ \prod_{h=1}^H P(Y_h=y_{ih} \vert X=t) \right\} \right],
    \end{equation*}
    to obtain estimate $\widehat{\boldsymbol{\pi}}^p$.
    Select for the first node of the decision tree the covariate $Z_{p^\star}$ for which the split yields the greatest log-likelihood; that is, $Z_p = Z_{p^\star}$ if $\ell_{2}(\boldsymbol{\pi}^p \vert \widehat{\boldsymbol{\phi}},S) > \ell_{2}(\boldsymbol{\pi}^{p^\prime} \vert \widehat{\boldsymbol{\phi}},S) \hspace{0.1cm} \forall \hspace{0.1cm} p^\prime \neq p$.
    The statistical significance of the selected node can be evaluated on the basis of the corresponding fitted slope parameters $\widehat{\gamma}_{2p^\star},\dots,\widehat{\gamma}_{Tp^\star}$ and their standard errors.
    \item For each covariate $Z_q \neq Z_{p^\star}$ and subsample, or branch, $S^{p_v^\star}$, fit \eqref{eq:covariates} with $\mathbf{Z} = Z_q$ to $S^{p_v^\star}$ conditionally on $\widehat{\boldsymbol{\phi}}$ by maximizing the pseudo log-likelihood
    \begin{equation*}
        \ell_{2}(\boldsymbol{\pi}^q \vert \widehat{\boldsymbol{\phi}},S^{p_v^\star}) = \sum_{i \in S^{p_v^\star}} \log \left[ \sum_{t=1}^T P(X=t \vert Z_q=v_i) \left\{ \prod_{h=1}^H P(Y_h=y_{ih} \vert X=t) \right\} \right],
    \end{equation*}
    to obtain estimate $\widehat{\boldsymbol{\pi}}^q$.
    For each branch, select the covariate $Z_{q^\star}$ for which the split yields the greatest log-likelihood; that is, $Z_q = Z_{q^\star}$ if $\ell_{2}(\boldsymbol{\pi}^q \vert \widehat{\boldsymbol{\phi}},S^{p_v^\star}) > \ell_{2}(\boldsymbol{\pi}^{q^\prime} \vert \widehat{\boldsymbol{\phi}},S^{p_v^\star}) \forall \hspace{0.1cm} q^\prime \neq q$.
    The statistical significance of the selected node can be evaluated on the basis of the corresponding fitted slope parameters $\widehat{\gamma}_{2q^\star},\dots,\widehat{\gamma}_{Tq^\star}$ and their standard errors.
    \item Analogously repeat step 3 for each remaining covariate $Z_w \neq Z_{p^\star}, Z_w \neq Z_{q^\star}, \dots$ and branch $S^{p_v^\star,q_v^\star,\dots}$.
    To avoid instability and overfitting, we propose splitting branches until the number of observed individuals in $S^{p_v^\star,q_v^\star,\dots}$ is smaller than some threshold.
    Without such a stopping rule, small changes in the data can lead to very different tree structures, and overfitting will be built-in by design; resulting in a maximum possible tree where each terminal node contains only 1 unique combination of values across the covariate space.
    \item Prune back the tree by, for each branch $S^{p_v^\star,q_v^\star,\dots,w_v^\star}$, omitting the current terminal split and re-pooling the corresponding subsamples $S^{p_v^\star,q_v^\star,\dots,w_0^\star}$ and $S^{p_v^\star,q_v^\star,\dots,w_1^\star}$.
    Continue pruning back the branch while this is locally optimal compared to keeping the split.
    We propose two alternative strategies based on different decision criteria: statistical significance or overall goodness-of-fit.
    For brevity of exposition of the algorithm, we describe these strategies in the subsequent section.
\end{enumerate}

The application of step-wise estimation, in which the measurement model is fitted separately before subsequently fitting the structural model, is fundamental to the methodological legitimacy of the proposed approach.
This is because adding and switching covariates, and fitting the model to different (sub)samples, can cause distortions to the measurement model if the full parameter space is fitted simultaneously.
Distortions of the measurement model is substantively equivalent to distortions of the class definitions, or interpretational confounding.
As such, different nodes of the decision tree could effectively correspond to covariate effects for different dependent variables.

The step-wise estimation approach presented above is the two-step estimator of \citet{bakkkuha2018}.
An alternative step-wise estimation approach is alternative bias-adjusted three-step estimator \citep{vermunt2010}.
This is less direct, as it involves an intermediate classification step between estimation of the measurement model and estimation of the structural model.
Nevertheless, the property avoidance of interpretational confounding is the same for both approaches.
Readers interested in differences between different step-wise estimation approaches are referred to \citet{vermunt2025}.

The proposed approach can be carried out in existing specialized latent class analysis software implementing step-wise estimation automatically or otherwise straightforwardly.
Examples of such softwares include the popular proprietary options \texttt{LatentGOLD} (free for academic use) and \texttt{Mplus}, and the popular open-source options \texttt{StepMix} in \texttt{Python} and \texttt{multilevLCA} in \texttt{R}.
After the next section, we illustrate the proposed approach by means of a real-data example using \texttt{multilevLCA}.
In the next section, we propose two alternative tree pruning strategies.

\section{Two tree pruning strategies}

\subsection{Statistical significance}

A first strategy for pruning back the decision tree and deciding when a split is locally optimal is based on statistical significance.
In this straightforward strategy, a branch is pruned back if the slope parameters for the current terminal split is non-significant.
As such, the strategy results in the maximum tree such that all the branches have significant terminal splits.

The statistical significance of a given split is evaluated on the basis of $T-1$ slope estimates.
As such, when $T>2$, more than one slope estimate is evaluated simultaneously.
It is possible that some of these estimates are statistically significant while other are not.
In such cases, the applied researcher may decide how many estimates must be statistically non-significant for a branch to be pruned back (for example, deciding that a branch is pruned back only if all $T>2$ estimates are statistically non-significant).

\subsection{Overall goodness-of-fit}

A second strategy is based on overall goodness-of-fit as measured by the Bayesian information criterion \citep[BIC;][]{schwarz1978}.
The BIC is a very popular model selection criterion with well-documented advantageous properties in standard and complex latent class analysis \citep[see e.g.][and references therein]{archnylund2026}.
Here, we distinguish overall goodness-of-fit for the full decision tree from local goodness-of-fit for a single node as calculated in the tree growing algorithm.
Overall goodness-of-fit can be computed on the basis of a constructed multinomial categorical covariate indicating the terminal node location of each unit.
We denote this constructed covariate by $\tau^{p_v^\star,q_v^\star,\dots,w_v^\star}$.

\eqref{eq:covariates} can be fitted with $\mathbf{Z}=\tau^{p_v^\star,q_v^\star,\dots,w_v^\star}$ to $S$ conditionally on the fitted measurement model $\widehat{\boldsymbol{\phi}}$.
This involves maximizing the pseudo log-likelihood

\begin{equation*}
    \ell_{2}(\boldsymbol{\pi}^{\tau^{p_v^\star,q_v^\star,\dots,w_v^\star}} \vert \widehat{\boldsymbol{\phi}},S) = \sum_{i=1}^N \log \left[ \sum_{t=1}^T P(X=t \vert \tau_i^{p_v^\star,q_v^\star,\dots,w_v^\star}) \left\{ \prod_{h=1}^H P(Y_h=y_{ih} \vert X=t) \right\} \right]
\end{equation*}
\\
to obtain estimate $\widehat{\boldsymbol{\pi}}^{\tau^{p_v^\star,q_v^\star,\dots,w_v^\star}}$.
Then, we can evaluate overall goodness-of-fit by computing the BIC as follows:

\begin{equation*}
    \text{BIC}^{\tau^{p_v^\star,q_v^\star,\dots,w_v^\star}} = \nu^{\tau^{p_v^\star,q_v^\star,\dots,w_v^\star}} \log(N) - 2\ell(\widehat{\boldsymbol{\pi}}^{\tau^{p_v^\star,q_v^\star,\dots,w_v^\star}},\widehat{\boldsymbol{\phi}},S),
\end{equation*}
\\
where $\nu^{\tau^{p_v^\star,q_v^\star,\dots,w_v^\star}}$ is the number of parameters estimated by the model.

This second tree pruning strategy can be described on the basis of three substeps.
In step 1, the BIC is computed for the current tree to be pruned; in step 2, one unique $\tau$ is constructed for each terminal split, such that each $\tau$ prunes back only its associated terminal split, and the BIC computed for each of these $\tau$s; in step 3, the model yielding the minimum BIC is selected.
These three steps are repeated for the selected tree until the step-1 tree yields the minimum BIC.

In applications where overfitting is a pregnant problem, this tree pruning strategy can be combined with cross-validation.
The following three steps describe how this can be operationalized:

\begin{enumerate}
    \item[(i)] Randomly divide the full sample $S$ into $G$ subsets, or folds, $S^g$.
    \item[(ii)] For each fold $S^g$ and constructed covariate $\tau^{p_v^\star,\dots}$, obtain a candidate model by maximizing $\ell_{2}(\boldsymbol{\pi}^{\tau^{p_v^\star,\dots}} \vert \widehat{\boldsymbol{\phi}},S^g)$ and compute overall goodness-of-fit $\text{BIC}_g^{\tau^{p_v^\star,\dots}}$ on the holdout set.
    \item[(iii)] For each constructed covariate $\tau^{p_v^\star,\dots}$, compute the mean overall goodness-of-fit across the $G$ folds, which we denote by $\overline{\text{BIC}}^{\tau^{p_v^\star,\dots}}$.
    Select the constructed covariate for which $\overline{\text{BIC}}^{\tau^{p_v^\star,\dots}}$ is the lowest.
\end{enumerate}
\
The subtree corresponding to the selected constructed covariate is taken as the final pruned tree.

Why do we evaluate goodness-of-fit by means of the BIC and not the log-likelihood as in the tree growing algorithm?
The reason is that the BIC adjusts fit for model complexity, which the log-likelihood does not.
In the tree growing algorithm, goodness-of-fit was evaluated locally for different models with the same number of parameters; as such, adjusting for fit would be redundant.
In the evaluating overall goodness-of-fit, the different models may have different numbers of parameters due to differences in the number of categories of $\tau$.
This warrants an adjustment for model complexity.

\section{Real-data illustration}

We illustrate the proposed approach for fitting decision trees to latent classes by means of an empirical example.
For ease of illustration, we apply the tree pruning strategy based on statistical significance.
While this empirical example may be smaller than the typical research context in which the proposed decision tree approach has clear advantages over the standard approach, we believe that this example is pedagogically more advantageous than a larger example involving more classes and covariates.
The illustration is based on public health data on 4546 American residents aged 20 years or above.
These data come from the 2017-2018 \emph{National Health and Nutrition Examination Survey} (NHANES) collected by the National Center for Health Statistics at the Centers for Disease Control and Prevention (CDC) in the United States \citep{johnson2013}.
The aim of the analysis is purely illustrative; we make no claims of contributing to any substantive literature.
All model estimation for the analysis is performed by means of the \texttt{R} package \texttt{multilevLCA} \citep{lyrvalletal2025}.

\begin{figure}[!ht]
    \centering
    \includegraphics[width=0.5\linewidth]{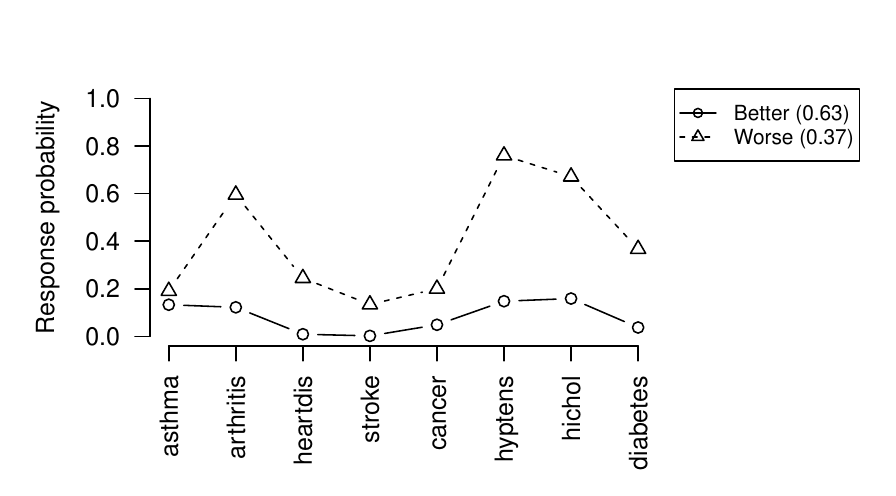}
    \caption{\footnotesize Fitted two-class measurement model of health classes.}
    \label{fig:classes}
\end{figure}

The measurement model is a dichotomous health classification with 8 binary indicators.
These indicate whether the respondent has ever been diagnosed with asthma (\emph{asthma}), arthritis (\emph{arthritis}), heart disease (\emph{heartdis}), stroke (\emph{stroke}), cancer or malignancy (\emph{cancer}), high blood pressure (\emph{hyptens}), or diabetes (\emph{diabetes}), or told their blood cholesterol was high (\emph{hichol}).
Figure~\ref{fig:classes} visualizes the fitted model, revealing a distinct partitioning of individuals into classes of better and worse health.
The class of individuals with worse health have consistently higher probabilities across all 8 indicators.
These have a probability above 50\% of having been diagnosed with arthritis, diagnosed with high blood pressure, or told their blood cholesterol was high.
The class of individuals with better health have probabilities below 20\% for each indicator.
The majority of individuals - 63\% - belong to the better-health class, and the remaining minority - 37\% - to the worse-health class.
Estimation of this standard latent class model corresponds to step 1 of the proposed algorithm.

Next, we add covariates to the model and carry out steps 2-5 of the algorithm to fit a decision tree to the health classes.
As covariates, we consider 4 binary survey items indicating whether the respondent is aged 60 years or above (\emph{age 60+}), recreationally active (\emph{active}), female (\emph{female}), or currently smokes cigarettes (\emph{smoker}).
The results are reported in Table~\ref{tab:tree1} and Table~\ref{tab:tree2}.
For the initial branch, or step 2, we fit the association between the classes and each of the covariates separately, while keeping the class definitions fixed to their plotted values in Figure~\ref{fig:classes}, by means of the two-step estimation approach \citep{bakkkuha2018}.
As shown in Table~\ref{tab:tree1}, we observe that the model with age as a covariate has the largest log-likelihood and the best fit to the data.
We thus perform the initial split on the basis of age.

\begin{table}[!ht]
    \centering
    \small
    \begin{tabular}{|rlrlrr|}
        \hline
         & \textbf{Branch} & \textbf{$N$} & \textbf{Covariate} & \textbf{$\ell$} & \textbf{$p$-value} \\
        \hline
        1. & Initial & 4546 & \emph{age 60+} & -14531.13 & 0.00 \\
         &  &  & \emph{active} & -15164.02 & 0.00 \\
         &  &  & \emph{female} & -15217.10 & 0.31 \\
         &  &  & \emph{smoker} & -15212.74 & 0.00 \\
        \hline
        2. & \emph{age 60+} $=0$ & 2848 & \emph{active} & -7198.59 & 0.00 \\
         &  &  & \emph{female} & -7208.16 & 0.46 \\
         &  &  & \emph{smoker} & -7206.50 & 0.05 \\
         &  &  &  &  &  \\
         & \emph{age 60+} $=1$ & 1698 & \emph{active} & -7319.42 & 0.01 \\
         &  &  & \emph{female} & -7319.61 & 0.02 \\
         &  &  & \emph{smoker} & -7321.28 & 0.11 \\
        \hline
        3. & \emph{age 60+} $=0$, \emph{active} $=0$ & 1353 & \emph{female} & -3631.80 & 0.89 \\
         &  &  & \emph{smoker} & -3631.16 & 0.26 \\
         &  &  &  &  &  \\
         & \emph{age 60+} $=0$, \emph{active} $=1$ & 1495 & \emph{female} & -3565.04 & 0.10 \\
         &  &  & \emph{smoker} & -3566.11 & 0.29 \\
         &  &  &  &  &  \\
         & \emph{age 60+} $=1$, \emph{active} $=0$ & 1075 & \emph{female} & -4740.09 & 0.00 \\
         &  &  & \emph{smoker} & -4743.35 & 0.07 \\
         &  &  &  &  &  \\
         & \emph{age 60+} $=1$, \emph{active} $=1$ & 623 & \emph{female} & -2574.50 & 0.82 \\
         &  &  & \emph{smoker} & -2574.35 & 0.55 \\
        \hline
    \end{tabular}
    \caption{\footnotesize Fitted first- to third-level branches of decision tree for two latent health classes.}
    \label{tab:tree1}
\end{table}

\begin{table}[!ht]
    \centering
    \small
    \begin{tabular}{|rlrlrr|}
        \hline
         & \textbf{Branch} & \textbf{$N$} & \textbf{Covariate} & \textbf{$\ell$} & \textbf{$p$-value} \\
        \hline
        4. & \emph{age 60+} $=0$, \emph{active} $=0$, \emph{smoker} $=0$ & 996 & \emph{female} & -2601.66 & 0.69 \\
         &  &  &  &  &  \\
         & \emph{age 60+} $=0$, \emph{active} $=0$, \emph{smoker} $=1$ & 357 & \emph{female} & -1029.41 & 0.95 \\
         &  &  &  &  &  \\
         & \emph{age 60+} $=0$, \emph{active} $=1$, \emph{female} $=0$ & 747 & \emph{smoker} & -1837.36 & 0.87 \\
         &  &  &  &  &  \\
         & \emph{age 60+} $=0$, \emph{active} $=1$, \emph{female} $=1$ & 748 & \emph{smoker} & -1726.89 & 0.22 \\
         &  &  &  &  &  \\
         & \emph{age 60+} $=1$, \emph{active} $=0$, \emph{female} $=0$ & 521 & \emph{smoker} & -2353.35 & 0.01 \\
         &  &  &  &  &  \\
         & \emph{age 60+} $=1$, \emph{active} $=0$, \emph{female} $=1$ & 554 & \emph{smoker} & -2383.50 & 0.50 \\
         &  &  &  &  &  \\
         & \emph{age 60+} $=1$, \emph{active} $=1$, \emph{smoker} $=0$ & 570 & \emph{female} & -2350.66 & 0.71 \\
         &  &  &  &  &  \\
         & \emph{age 60+} $=1$, \emph{active} $=1$, \emph{smoker} $=1$ & 53 & \emph{female} & -223.61 & 0.90 \\
        \hline
    \end{tabular}
    \caption{\footnotesize Fitted fourth-level branches of decision tree for two latent health classes.}
    \label{tab:tree2}
\end{table}

For the second-level branches, or step 3 of the algorithm, we fit the structural model with activity status, the structural model with sex, and the structural model with smoking status to each of two subsamples: the subsample of respondents that are younger than 60 years (\emph{age 60+} $=0$ in Table~\ref{tab:tree1}), and the subsample of respondents that are 60 years or older.
We observe that the structural model with activity status has the best fit to each of these subsamples.
The third-level branches, or step 4, involve fitting the two structural models with sex and smoking status to each of the four subsamples combining the two age categories with the two activity statuses.
The results show that the model with sex has the better fit to the subsample of younger and active people (\emph{age 60+} $=0$, \emph{active} $=1$ in Table~\ref{tab:tree1}) and the subsample of older and inactive people, while the model with smoking status has the better fit to the subsample of younger and inactive people and the subsample of older and active people.

For the fourth-level branches, or step 5, we fit eight structural models with a single covariate to each of the subsamples combining the two age categories, the two activity statutes, and either of the two sexes or smoking statuses, as shown in Table~\ref{tab:tree2}.
Finally, we prune back the tree based on statistical significance, such that a branch is pruned while its terminal split is statistically non-significant at the 5\%-level.
We can observe that a single terminal node in the full tree is statistically significant, namely the node corresponding to the structural model with smoking status fitted to the subsample of older, inactive, male individuals (\emph{age 60+} $=1$, \emph{active} $=0$, \emph{female} $=0$ in Table~\ref{tab:tree1}).
This branch is retained while the remaining seven branches are pruned back.

Moving up the tree and inspecting the results for the three third-level branches in Table~\ref{tab:tree1}, we find that none of the three new terminal nodes (i.e. all branches except \emph{age 60+} $=1$, \emph{active} $=0$ which was retained at the fourth level) is statistically significant.
We thus move further back and inspect the results for the second-level branches.
Here we can observe that the new terminal node in the subsample of younger individuals is statistically significant.
As such, all terminal nodes of the current tree are statistically significant, the pruning halts, and we take the current tree as final.

Before displaying and inspecting the final tree, we make a note on sample size.
At the fourth branch level, we fitted a structural model to a small subsample of 53 individuals (\emph{age 60+} $=1$, \emph{active} $=1$, \emph{female} $=1$ in Table~\ref{tab:tree2}).
While we pruned back this branch because of statistical non-significance ($p=0.90 > 0.05$), we could arguably have omitted fitting it in the first place because of its small sample size if we had applied a minimum threshold of, say, 100.
However, in this setting, with two latent classes and a single binary covariate, we can reasonably consider a sample size of 53 to be sufficient.

\begin{figure}[!ht]
    \centering
    \includegraphics[width=0.5\linewidth]{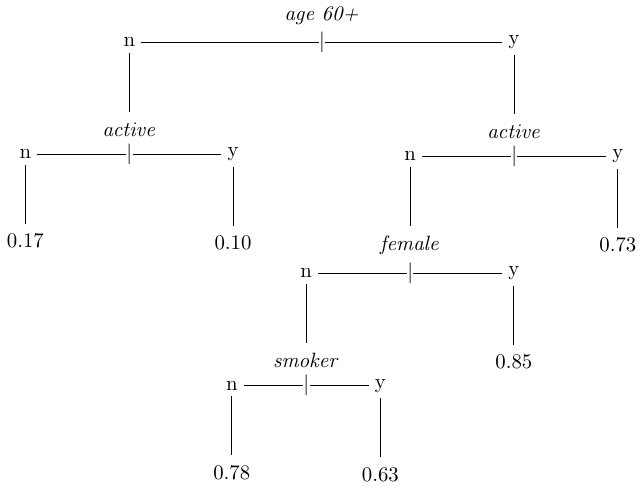}
    \caption{\footnotesize Fitted (final) decision tree for two latent health classes.}
    \label{fig:tree}
\end{figure}

Figure~\ref{fig:tree} illustrates the final decision tree.
To enhance readability, we use here \emph{n} for \emph{no} and \emph{y} for \emph{yes} where in Table~\ref{tab:tree1} and Table~\ref{tab:tree2} we used 0 and 1, respectively.
Because our dependent variable has two latent classes, we display the probabilities for a single class, namely the worse-health class.
The table can be read as follows.
Individuals aged below 60 years have a low probability of belonging to the worse-health class, yet this probability varies with activity status: younger individuals that are active (10\%) have a lower probability compared to younger individuals that are inactive (17\%).
Among older individuals, the probability of belonging to the worse-health class is high yet variable.
Those that are active have a 73\% probability.
This is lower than the probability for inactive females (85\%) which have the highest probability overall.
Among inactive males, smokers have a 78\% probability, and non-smokers a 63\% probability, of belonging to the worse-health class.\footnote{
We recall that the smoking covariate indicates present - not historical - smoking.
The result that smoking is associated with a smaller prevalence of worse health among older inactive males may appear odd, but can have logical reasons, for example, survivor bias, or that these individuals may see no reason to give up smoking if their health is proven to be fine.
}

As such, this example illustrates the potential of our proposed methodology for integrating decision tree modeling into latent class analysis to uncover complex interactions among the covariates.

\section{Concluding remarks}

We have proposed an approach for modeling complexly interacting covariate effects by means of a decision tree, and illustrated it by means of an applied latent class analysis of painter types.
This offers an alternative to standard (multinomial) logistic regression for contexts with complexity of interpretation of linear interaction effects.
The innovative decision tree structure for covariate effects is highly interpretable, leading the eye through combinations of covariates to a final conditional classification.
Our novel methodology does not require any non-traditional assumptions, is based on well-established routines of model estimation, and can be readily implemented in existing software.

While this work has focused on binary covariates, the routine can be easily extended to multinomial and continuous covariates.
This involves dividing each non-binary covariate into a set of split points, and fitting the structural model for each covariate-splitpoint combination in the same manner as for binary covariates.
The routine enumerates the number of analytical sub-steps, but retains the statistical and conceptual properties of the proposed approach.
One important extension of this work would be to extend the routine to the general case allowing for multinomial and continuous covariates, by developing more advanced within-predictor splitting procedures.

The closest methodological relative to the proposed approach is a hybrid method combining the CHAID algorithm and latent class analysis \citep{magidsonvermunt2005}.
This involves first computing posterior class membership probabilities, and then treating them as an observed dependent variable while constructing a tree.
In contrast to this approach, the proposed approach is fully compatible with the benchmark way of specifying latent class models: taking the covariates as affecting class membership probability, as opposed to affecting posterior class membership probability.

One limitation of the current approach is its reliance on local optima for the splits.
This arises from its recursive nature, wherein the decision tree is built based on local optimality at the current step and not a potential future step.
This limitation is common to decision tree modeling generally, not a particular feature of its integration into latent class analysis.

A promising direction of future research is the development and evaluation of alternative tree pruning strategies.
In this paper, we have proposed two strategies, based on statistical significance and goodness-of-fit.
While these may be taken as natural strategies, we are not claiming that these be the only natural strategies.
Goodness-of-fit can be evaluated by means of different measures than the BIC, for instance the exact (non-asymptotic) integrated complete likelihood \citep[or ICL;][]{biernackiceleuxgovaert2010}, which is used for model selection in other latent class analysis methodology \citep{lyrvalletal2026}.
One may also consider the ``mix'' of the two in the form of the ICLbic criterion \citep{biernackiceleuxgovaert2002}.
Different evaluation measures than goodness-of-fit can be considered, such as prediction quality measures like the entropy-based R$^2$.
The methodological literature on decision tree modeling within latent class analysis, which we aim to establish in this paper, can benefit from more rigorous comparisons of these and future strategies, and, more generally, from scholarly discussion on this topic.

The ultimate aim of the present proposal of an integration of decision trees into latent class modeling with covariates is to initiate a novel research line in the latent class analysis methodological literature, and to facilitate the advancement of applied research in diverse disciplines with a novel tool for extracting new insights from empirical data.

\clearpage

\bibliography{lyrvall-clouth-2026}

\end{document}